\documentclass[aps,prb,longbibliography,showpacs,twocolumn,superscriptaddress]{revtex4-2}
\usepackage{amsmath,amssymb,amsfonts,bm}
\usepackage{booktabs}
\usepackage{braket}
\usepackage{graphicx}
\usepackage{epstopdf}
\usepackage{dcolumn}
\usepackage{mathrsfs}
\usepackage{bbold}
\usepackage{dsfont}
\usepackage{float}

\usepackage{tgtermes}
\usepackage{multirow}
\usepackage[colorlinks=true,linkcolor=blue,citecolor=blue, urlcolor=blue,bookmarks=false]{hyperref}

\begin{document}
	
\title{Hybrid-order topology in two-dimensional nonsymmorphic antiferromagnets} 
\author{Wei Xiong}
\thanks{These authors contributed equally to this work.}
\affiliation{Department of Physics and Chongqing Key Laboratory for Strongly Coupled Physics, Chongqing University, Chongqing 400044,  China}

\author{Zi-Ming Wang}
\thanks{These authors contributed equally to this work.}
\affiliation{Department of Physics and Chongqing Key Laboratory for Strongly Coupled Physics, Chongqing University, Chongqing 400044,  China}

\author{Xin-Mei Wei}
\affiliation{Department of Physics and Chongqing Key Laboratory for Strongly Coupled Physics, Chongqing University, Chongqing 400044,  China}

\author{Rui Wang}
\affiliation{Department of Physics and Chongqing Key Laboratory for Strongly Coupled Physics, Chongqing University, Chongqing 400044, China}
\affiliation{Center of Quantum Materials and Devices, Chongqing University, Chongqing 400044, China}

\author{Dong-Hui Xu}
\email{donghuixu@cqu.edu.cn}
\affiliation{Department of Physics and Chongqing Key Laboratory for Strongly Coupled Physics, Chongqing University, Chongqing 400044, China}
\affiliation{Center of Quantum Materials and Devices, Chongqing University, Chongqing 400044, China}


\begin{abstract}
We theoretically demonstrate hybrid-order topology in a two-dimensional nonsymmorphic antiferromagnet. Utilizing a generic antiferromagnetic Dirac model with a symmetry-allowed, momentum-dependent spin-density-wave (SDW) mass, we show that a single bulk insulating phase exhibits distinct topological boundary manifestations governed solely by the termination geometry. For screw-compatible edges, nonsymmorphic screw symmetry protects gapless first-order edge states. In contrast, for a $45^\circ$ diamond-shaped termination, the screw symmetry is broken at the boundary, resulting in gapped edges. However, the finite geometry still preserves magnetic mirror symmetries $\mathcal{M}_x\mathcal{T}$ and $\mathcal{M}_y\mathcal{T}$, which enforce an alternating pattern of edge masses, thereby binding zero-dimensional corner states. This second-order phase is characterized by a quantized quadrupole moment, with corner states pinned to zero energy by the chiral symmetry. We further demonstrate that explicit lattice perturbations can selectively gap the first-order edge modes while robustly preserving the corner states. Our work establishes a symmetry-based route to a termination-controlled duality between first- and second-order topology in magnetic nonsymmorphic systems.
\end{abstract}
\maketitle
	
\section{Introduction} Topological insulators (TIs) represent a paradigm of symmetry-protected matter, typically characterized by time-reversal symmetry (TRS) and a $\mathbb{Z}_2$ invariant~\cite{RevModPhys.82.3045,RevModPhys.83.1057}. Historically, this framework appeared to exclude magnetic systems, as magnetic order inherently breaks TRS. However, this paradigm expanded significantly with the proposal of antiferromagnetic~(AFM) topological insulators, wherein topology is protected by an effective antiunitary symmetry combining time reversal with a half-lattice translation~\cite{PhysRevB.81.245209,vsmejkal2018topological}. Related work further clarified the topological classification of commensurate antiferromagnets~\cite{PhysRevB.88.085406}. This concept was later realized in the MnBi$_2$Te$_4$ family, which conclusively exhibited intrinsic AFM topological behavior both theoretically and experimentally~\cite{PhysRevLett.122.206401,PhysRevLett.122.107202,otrokov2019prediction,Gong2019IntrinsicMTI,Li2024MnBi2Te4Review}. In two dimensions, the AFM quantum spin Hall effect was subsequently proposed in FeSe/SrTiO$_3$ thin films~\cite{Wang2016Topological} and in XMnY (X = Sr, Ba; Y = Sn, Pb)~\cite{PhysRevLett.124.066401}.

Conventionally, both non-magnetic and magnetic TIs adhere to the standard bulk-boundary correspondence, where a $D$-dimensional bulk hosts $(D-1)$-dimensional boundary states. Recently, this principle has been generalized to higher-order topological insulators (HOTIs). In these systems, a $d$-th order topological phase in $D$ dimensions manifests localized boundary states on lower-dimensional, $(D-d)$-dimensional boundaries~\cite{PhysRevLett.110.046404,benalcazar2017quantized,PhysRevLett.119.246402,schindler2018higher_order,xie2021higher_order_band,Wieder2022_TopologicalMaterialsDiscovery,PhysRevLett.120.026801,PhysRevLett.123.177001,PhysRevLett.124.036803,PhysRevB.102.241102,PhysRevLett.123.256402,PhysRevLett.123.216803,lee2020two}. For example, a two-dimensional second-order topological insulator hosts zero-dimensional corner states rather than conventional one-dimensional edge modes. By extending the bulk-boundary correspondence, higher-order topology provides novel mechanisms for bulk invariants to manifest in lower-dimensional boundary responses. Furthermore, recent studies have proposed hybrid topological phases, where first-order and higher-order topological states coexist simultaneously within the same physical system~\cite{PhysRevB.99.125149, PhysRevB.102.041122,PhysRevB.111.195107,l1n5-1jsm,zpcc-59kv,PhysRevLett.126.156801}.

Higher-order topology in antiferromagnets has been explored across multiple dimensions. In three-dimensional AFM materials, second-order topological phases with hinge states were proposed in the axion-insulator EuIn$_2$As$_2$~\cite{PhysRevLett.122.256402,PhysRevResearch.2.043274} and in MnBi$_{2n}$Te$_{3n+1}$ systems~\cite{PhysRevLett.124.136407}. In two dimensions, AFM second-order topological insulators with corner states have been investigated from both model and materials perspectives~\cite{Mu2022_AFMSecondOrderSOTI,Luo2022Fragile,Zhan2024_AFMSecondOrderBandTopology}. While these studies demonstrate that antiferromagnetism and higher-order topology are inherently compatible, first-order and higher-order AFM topological phases have largely been treated as mutually exclusive phenomena~\cite{bernevig2022progress}.

In this work, we address an open question: Can a single AFM bulk phase exhibit a duality of boundary topologies—switching between first-order topological boundary states and higher-order topological boundary states—dictated solely by the symmetries preserved at the termination? We demonstrate that this duality is indeed realizable in generic nonsymmorphic antiferromagnets. In our proposed model, first-order edge states are protected by screw symmetries, while second-order corner states are stabilized by magnetic mirror symmetries. A crucial theoretical ingredient is the inclusion of a symmetry-allowed, momentum-dependent spin-density-wave (SDW) mass, motivated by unconventional SDW theory~\cite{PhysRevLett.81.3723}. This term reshapes the bulk Dirac masses in momentum space and naturally generates the boundary mass inversion required for higher-order topology. We further show that boundary orientation and explicit symmetry-breaking perturbations can selectively gap the first-order edge states while robustly preserving the corner states. The resulting second-order topological phase is rigorously characterized by a quantized quadrupole moment. Ultimately, our findings establish a comprehensive framework for hybrid-order topology in antiferromagnets, revealing that the observable manifestation of topology is fundamentally governed by the interplay between magnetic symmetry sectors and boundary geometry.

\section{Model and symmetry analysis} 
\begin{figure}[t!]
    \centering
    \includegraphics[width=\columnwidth]{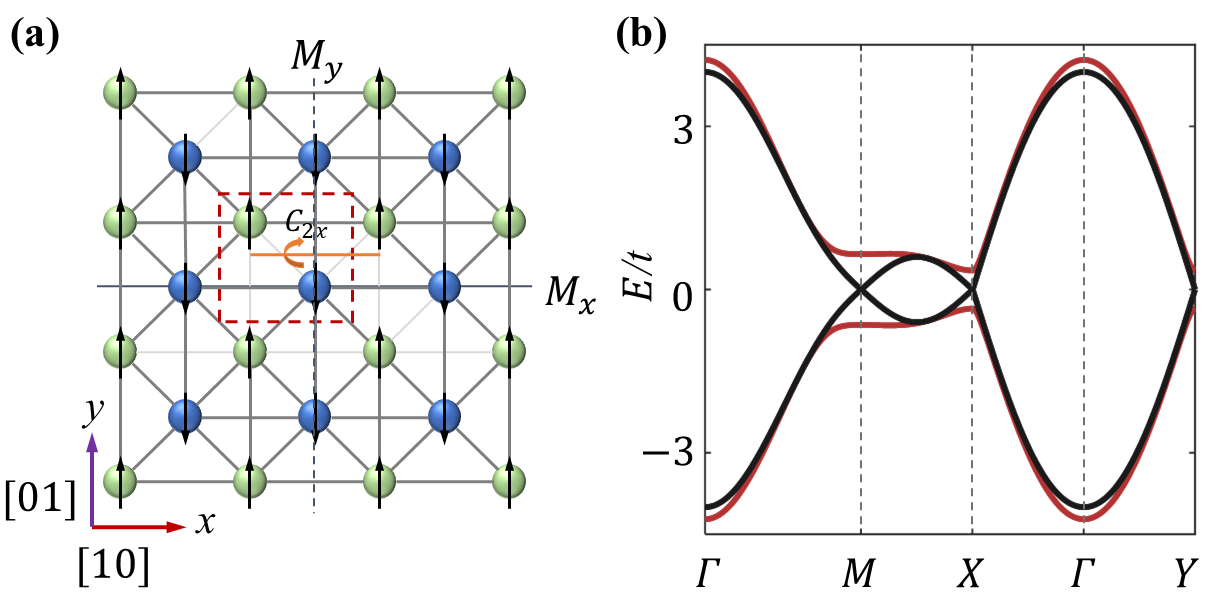}
    \caption{(a) Schematic of the lattice structure. The two sublattices are shown as green and blue filled circles, and the staggered out-of-plane Néel order is depicted by up and down arrows. (b) Bulk electronic band structure of the nonmagnetic phase (black lines), showing symmetry-enforced Dirac points at the high-symmetry momenta $X$, $Y$, and $M$. The red lines are obtained after turning on the SDW mass in Eq.~\ref{eq:mass}, which gaps the Dirac crossings and produces the AFM insulating phase. Unless otherwise specified, the parameters are $t=1.0$, $t_{\rm soc}=0.3$, and $\Delta_0=-0.35$, $\Delta_{1}=-0.25$. }
    \label{fig1}
\end{figure}

We consider a two-dimensional square lattice with two sublattices per unit cell and an out-of-plane N\'eel-type AFM SDW order, as illustrated in Fig.~\ref{fig1}(a). The model is intended as a minimal symmetry-based description of a nonsymmorphic AFM Dirac system. Its lattice and spin-orbit structure are nevertheless closely related to previously proposed AFM Dirac models in nonsymmorphic settings~\cite{PhysRevLett.118.106402,PhysRevB.95.115138}. 

In momentum space,  the Hamiltonian in the basis $\Psi_{\bm{k}} = (c_{\bm{k} \text{A} \uparrow}, c_{\bm{k} \text{A} \downarrow}, c_{\bm{k} \text{B} \uparrow}, c_{\bm{k} \text{B} \downarrow})^T$ reads:
\begin{equation}\label{eq:HK}
\begin{split}
   H = & t\cos \frac{k_x}{2}\cos \frac{k_y}{2}\tau_x +t_{\rm soc}\tau_z(\sigma_y\sin k_x-\sigma_x\sin k_y)
   \\ &  + M(\bm k)\tau_z\sigma_z,
\end{split}
\end{equation}
where the Pauli matrices $\tau_i$ and $\sigma_i$ act on the sublattice and spin degrees of freedom, respectively. The first term describes inter-sublattice hopping generated by the nonsymmorphic half translation. The second term represents a Rashba-type spin-orbit coupling allowed by the broken local inversion symmetry. The final term is the staggered exchange field associated with the AFM SDW order.

The nonmagnetic parent phase is described by the paramagnetic space group $P4/nmm1'$. In this phase, the system preserves inversion symmetry $\mathcal{P}=\tau_x$, TRS $\mathcal{T}=i \sigma_y\mathcal{K}$, and two orthogonal screw rotations, $\mathcal{S}_x=\{C_{2x}|\tfrac{1}{2}0\}=ie^{-ik_x/2}\tau_x\sigma_x$ and $\mathcal{S}_y=\{C_{2y}|0\tfrac{1}{2}\}=ie^{-ik_y/2}\tau_x\sigma_y$. In this phase, the system is a two-dimensional Dirac semimetal that hosts symmetry-enforced Dirac crossings at $X$, $Y$, and $M$~\cite{PhysRevLett.115.126803}, as shown in Fig.~\ref{fig1}(b). These crossings are stabilized by the combined action of $\mathcal{PT}$ and the nonsymmorphic screw symmetries, which generate projective representations on the Brillouin-zone boundary and prevent hybridization between states carrying different symmetry eigenvalues. 

Upon introducing the N\'eel order, the magnetic space group is reduced to $P4/n'm'm'$. Although $\mathcal{P}$ and $\mathcal{T}$ are individually broken by the antiferromagnetic ordering, their product $\mathcal{PT}$ is preserved, along with the magnetic mirror symmetries $\mathcal{M}_x\mathcal{T}=\sigma_z\mathcal{K}$ and $\mathcal{M}_y\mathcal{T}=\mathcal{K}$. Once the AFM SDW order is introduced, $\mathcal{PT}$ continues to guarantee a twofold Kramers-like degeneracy at each momentum, but it no longer protects the Dirac crossings. The SDW order therefore opens a full bulk gap. In addition to the crystallographic and magnetic symmetries discussed, the minimal model also satisfies a chiral symmetry $\mathcal{C} = \tau_y$, which enforces an energy spectrum that is exactly symmetric around zero energy.

To capture the leading finite-range correction to the AFM order, we define the momentum-dependent mass:
\begin{equation}
M(\bm{k})=\Delta_0+2\Delta_1(\cos k_x+\cos k_y),
\label{eq:mass}
\end{equation}
where $\Delta_0$ is the onsite N\'eel component and $\Delta_1$ is an extended-$s$-wave bond-SDW term~\cite{PhysRevLett.81.3723}. Since both terms transform as the fully symmetric $A_{1g}$ representation, $M(\bm{k})\tau_z\sigma_z$ introduces no symmetry breaking beyond the primary AFM order. 

This momentum dependence fundamentally dictates the topological character of the bulk. While the uniform term $\Delta_{0}$ independently opens a trivial global gap, the momentum-dependent term $\Delta_{1}$ modulates the local Dirac masses across the Brillouin zone. The topological regime is controlled by the relative sign of the masses at $M$ and $X/Y$. Specifically, a topological band inversion occurs at $M$ when $\Delta_0(\Delta_0 - 4\Delta_1)<0$.

Importantly, the screw symmetries remain physically meaningful only for boundaries parallel to their respective axes, because each contains a fractional translation. As a result, the boundary manifestation of the AFM topological phase depends strongly on the crystallographic termination.

\section{Boundary-selective first-order topology} 
\begin{figure}[t!]
    \centering
    \includegraphics[width=\columnwidth]{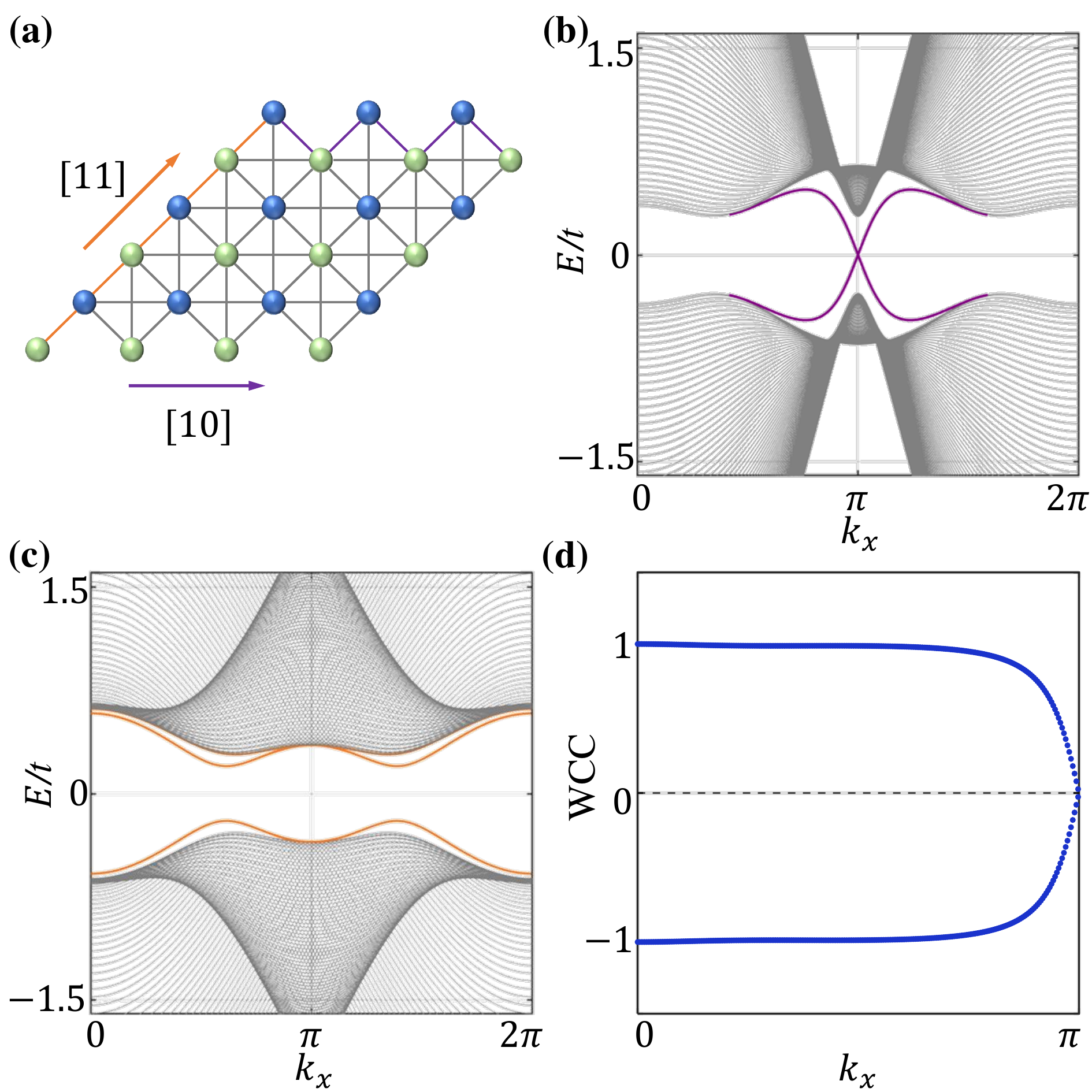}
    \caption{{Boundary-selective first-order topology. (a) Real-space geometry of the nanoribbon. The purple and yellow arrows represent the periodic boundary directions for (b) and (c), respectively. (b) Edge spectrum for the screw-preserving ribbon, showing symmetry-protected gapless edge modes. (c) Edge spectrum for a ribbon with boundary orientation incompatible with the screw symmetries, showing that the edge modes become gapped. (d) Wannier charge-center evolution along $k_x$, exhibiting a nontrivial flow consistent with the $\mathbb{Z}_2$ boundary topology of the screw-preserving edge.}}
    \label{fig2}
\end{figure}

Because a screw symmetry contains a fractional translation, it can be preserved only on boundaries parallel to the corresponding screw axis. We first consider a nanoribbon cleaved along the $x$ direction, i.e., with a $[10]$ edge, as shown in Fig.~\ref{fig2}(a). This boundary preserves the screw symmetry $\mathcal{S}_x=\{C_{2x}|\tfrac{1}{2}0\}$. In the resulting one-dimensional edge theory, $\mathcal{S}_x$ forbids any constant Dirac mass term, so the boundary spectrum remains gapless. The dispersive states in Fig.~\ref{fig2}(b) are therefore screw-protected first-order edge modes rather than accidental boundary resonances.

The Wannier charge-center (WCC) evolution shown in Fig.~\ref{fig2}(d) is consistent with this nontrivial crystalline topology. More specifically, the nontrivial WCC flow indicates a bulk phase whose gapless boundary manifestation requires a termination compatible with the corresponding screw symmetry sector. This interpretation is fully consistent with the observation that only screw-preserving boundaries support metallic edge states.

The situation changes qualitatively for a diagonal $[11]$ termination. Such an edge is microscopically stepped and necessarily mixes the two sublattices, thereby breaking the fractional translation entering the screw operations. Once the boundary screw symmetry is lost, a constant edge Dirac mass becomes symmetry allowed, and the edge spectrum is fully gapped, as shown in Fig.~\ref{fig2}(c). The same bulk AFM phase thus exhibits a boundary-selective first-order response: screw-compatible terminations host gapless edge states, whereas screw-incompatible terminations do not.

\section{Second-order topology in the diamond geometry} 
\begin{figure}[t!]
    \centering
    \includegraphics[width=\columnwidth]{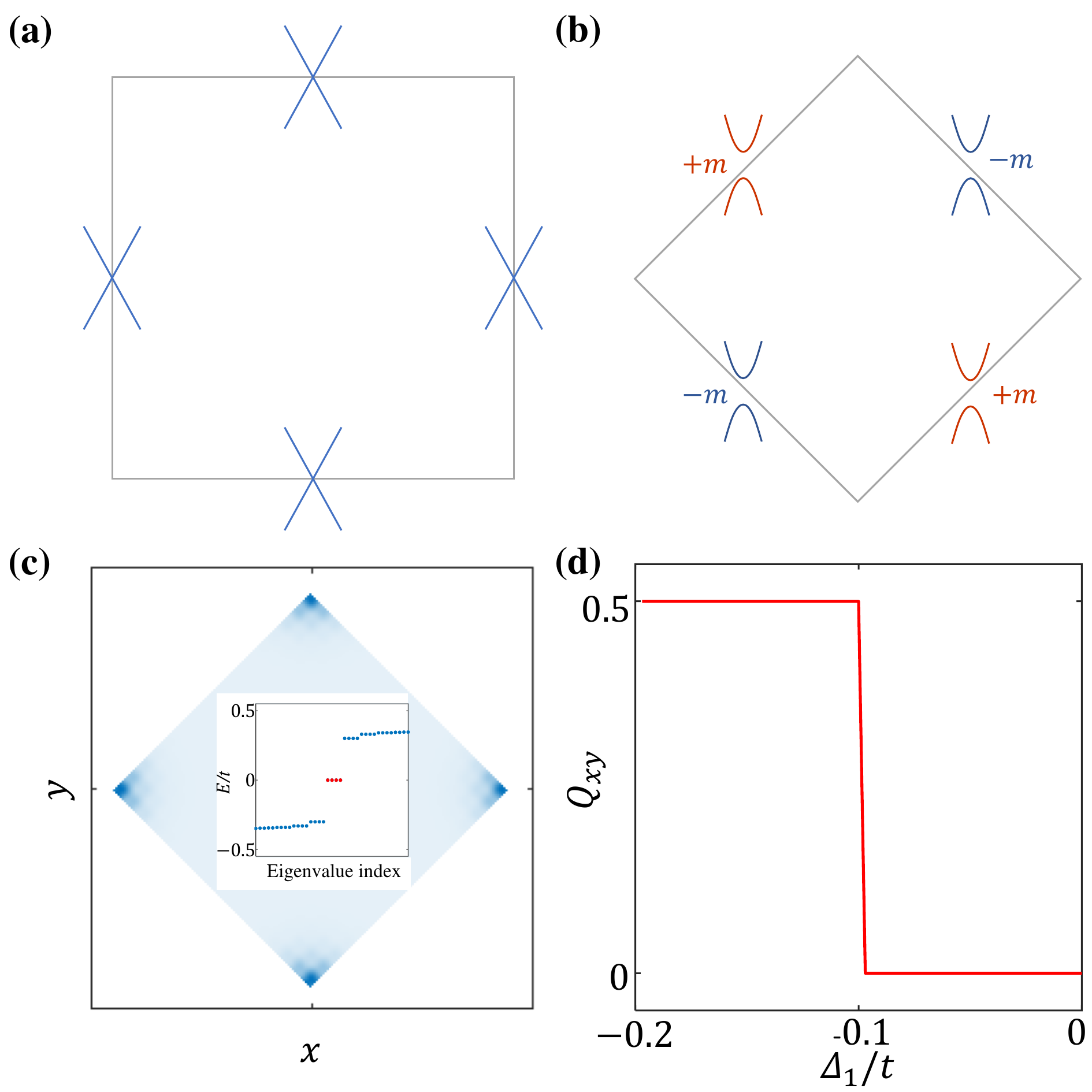}
    \caption{{Second-order topological phase in the diamond geometry. (a) Schematic of a square-like sample whose screw-preserving edges support gapless first-order boundary modes. (b) Diamond geometry with gapped edges; $+m$ and $-m$ denote the alternating Dirac masses enforced on neighboring edges by the magnetic mirror symmetries. (c) Center: energy spectrum of the fully finite sample corresponding to (b), showing four in-gap states pinned near zero energy (red). Surrounding panels: spatial probability distributions of the four in-gap states, demonstrating strong localization at the corners and negligible weight along the edges or in the bulk. (d) Real-space quadrupole moment $Q_{xy}$ as a function of $\Delta_1$, revealing the topological phase transition into the higher-order phase.} }
    \label{fig3}
\end{figure}

A termination with fully gapped edges need not be topologically trivial. To determine the fate of the AFM phase under screw-breaking boundaries, we consider a fully finite sample with a diamond geometry bounded by $[11]$ and $[1\bar{1}]$ edges. In contrast to a square geometry [Fig.~\ref{fig3}(a)], whose boundaries can preserve the screw symmetries and therefore support gapless edge states, the diamond geometry breaks boundary-preserving screw symmetry on every edge. Consequently, all four edges become gapped.

Remarkably, the gapped diamond is nevertheless topological. Although the screw symmetries are absent at the boundary, the full sample still preserves the magnetic mirror symmetries $\mathcal{M}_x\mathcal{T}$ and $\mathcal{M}_y\mathcal{T}$. These antiunitary symmetries do not protect gapless edges directly. Instead, they constrain the signs of the allowed Dirac masses on adjacent edges. The result is an alternating edge-mass pattern, $+m$, $-m$, $+m$, $-m$, around the perimeter, as illustrated in Fig.~\ref{fig3}(b). Each corner is therefore a domain wall between neighboring edges with opposite Dirac masses. By the Jackiw--Rebbi mechanism, such domain walls bind zero-energy states.

This analytical picture is verified by numerical diagonalization in Fig.~\ref{fig3}(c). The finite-size spectrum reveals four in-gap states that are well separated from both the bulk bands and the gapped edge continuum. Protected by the chiral symmetry $\tau_y$, these states are pinned exactly to zero energy.  The corresponding real-space wave-function profiles demonstrate that these states are sharply localized at the four corners, with negligible weight in the interior or along the edges. This distinct spatial configuration confirms that the corner modes are not finite-size remnants of first-order edge states, but genuine codimension-two boundary excitations.

To further characterize the higher-order phase, we compute the real-space quadrupole moment
\begin{equation}
Q_{xy}=\mathrm{Im}\left[\frac{1}{2\pi}\operatorname{Tr}\log\left(U^\dagger e^{2\pi i \hat q_{xy}} U\right)\right]-n_{\mathrm{al}},
\label{eq:quadrupole}
\end{equation}
where $U$ is the matrix of occupied eigenstates, $\hat q_{xy}=\hat x\hat y/(L_xL_y)$, and $n_{\mathrm{al}}$ subtracts the atomic-limit contribution. In the present symmetry setting, $\mathcal{M}_x\mathcal{T}$ and $\mathcal{M}_y\mathcal{T}$ quantize $Q_{xy}$ modulo $1$, provided the dipole polarizations vanish. As shown in Fig.~\ref{fig3}(d), the discontinuous change of $Q_{xy}$ as $\Delta_1$ is tuned signals the transition into the second-order phase. Physically, this shows that the extended SDW component is not a minor correction to the N\'eel mass, but the control parameter that reshapes the bulk mass structure and drives the boundary mass inversion.

The stark contrast between the square and diamond geometries isolates the essence of this hybrid-order phenomenon. The bulk Hamiltonian remains identical, yet the boundary topology switches qualitatively because the respective geometries preserve orthogonal symmetry sectors. The system transitions from a first-order to a second-order topological manifestation purely through geometric termination, without undergoing a bulk phase transition.

\section{Edge theory and perturbative robustness} The numerical results can be understood analytically from a low-energy theory around the $M$ point, $\bm{k}=(\pi,\pi)+\bm{q}$, where the momentum-dependent SDW term most strongly reshapes the bulk gap. To leading order we obtain
\begin{equation}
H_{M}(\bm{q})=
\left[\tilde{\Delta}+\Delta_1(q_x^2+q_y^2)\right]\tau_z\sigma_z
+t_{\rm soc}(q_y\tau_z\sigma_x-q_x\tau_z\sigma_y),
\label{eq:HM}
\end{equation}
with $\tilde{\Delta}=\Delta_0-4\Delta_1$. The subleading term $(t/4)q_xq_y\tau_x$ is omitted because it does not modify the leading boundary mass structure or the topology of the edge theory. Equation~(\ref{eq:HM}) makes the role of $\Delta_1$ particularly transparent: it renormalizes the Dirac mass at $M$ and determines the sign of the low energy gap governing the boundary theory.

We first consider a $(10)$ edge modeled by the half-plane $x\ge 0$. Replacing $q_x\rightarrow -i\partial_x$, we separate the Hamiltonian into a transverse part and a longitudinal perturbation
\begin{subequations}
    \begin{align}
        H_0 & = (\tilde{\Delta}-\Delta_1\partial_x^2)\tau_z\sigma_z+it_{\rm soc}\partial_x\tau_z\sigma_y , \\
        H_{\text{p}} & = t_{\rm soc}q_y\tau_z\sigma_x .
    \end{align}
\end{subequations}
Solving $H_0\psi(x)=0$ yields a two-dimensional edge-state subspace. Projecting $H_{\text{p}}$ into this subspace gives the effective edge Hamiltonian
\begin{equation}
    H_{{\rm eff},[10]}=v q_y\,{\eta}_z ,
\end{equation}
where ${\eta}_z$ act in the projected edge basis. The absence of a constant mass term is the essential point: for the $[10]$ edge, the preserved screw symmetry forbids such a term, so the boundary spectrum remains gapless. The same argument applies to the $[01]$ edge. This explains why the ribbon in Fig.~\ref{fig2}(b) exhibits robust first-order edge modes.

The diagonal $[11]$ and $[1\bar{1}]$ edges of the diamond sample behave differently. Because these edges do not preserve the local screw symmetries, a constant mass term is now symmetry allowed in the projected boundary theory~\cite{PhysRevX.12.011030},
\begin{equation}
H_{{\rm eff},{\rm edge}}=v q_{\parallel}\eta_z+m_{\rm e}\eta_x.
\label{eq:edge_mass}
\end{equation}

Each diagonal edge is therefore individually gapped. However, the edge masses are not independent. After projection to the edge subspace, the magnetic mirrors $\mathcal{M}_x\mathcal{T}$ and $\mathcal{M}_y\mathcal{T}$ relate neighboring diagonal edges and reverse the sign of the corresponding mass term. The four edges of the diamond must thus carry an alternating mass pattern $\{+m,-m,+m,-m\}$, up to an overall sign. The corner modes observed in Fig.~\ref{fig3}(c) are precisely the Jackiw--Rebbi bound states associated with these mass domain walls.

It is important to distinguish the origin of the corner states from their exact spectral pinning. The existence of the corner modes follows from the magnetic-mirror-enforced inversion of the edge masses, whereas their precise zero-energy pinning is additionally stabilized by the chiral symmetry. In this sense, the second-order topology is protected by the edge-mass inversion pattern, while the exact midgap location of the corner states reflects the additional spectral constraint of the minimal lattice model.

\begin{figure}[t!]
    \centering
    \includegraphics[width=\columnwidth]{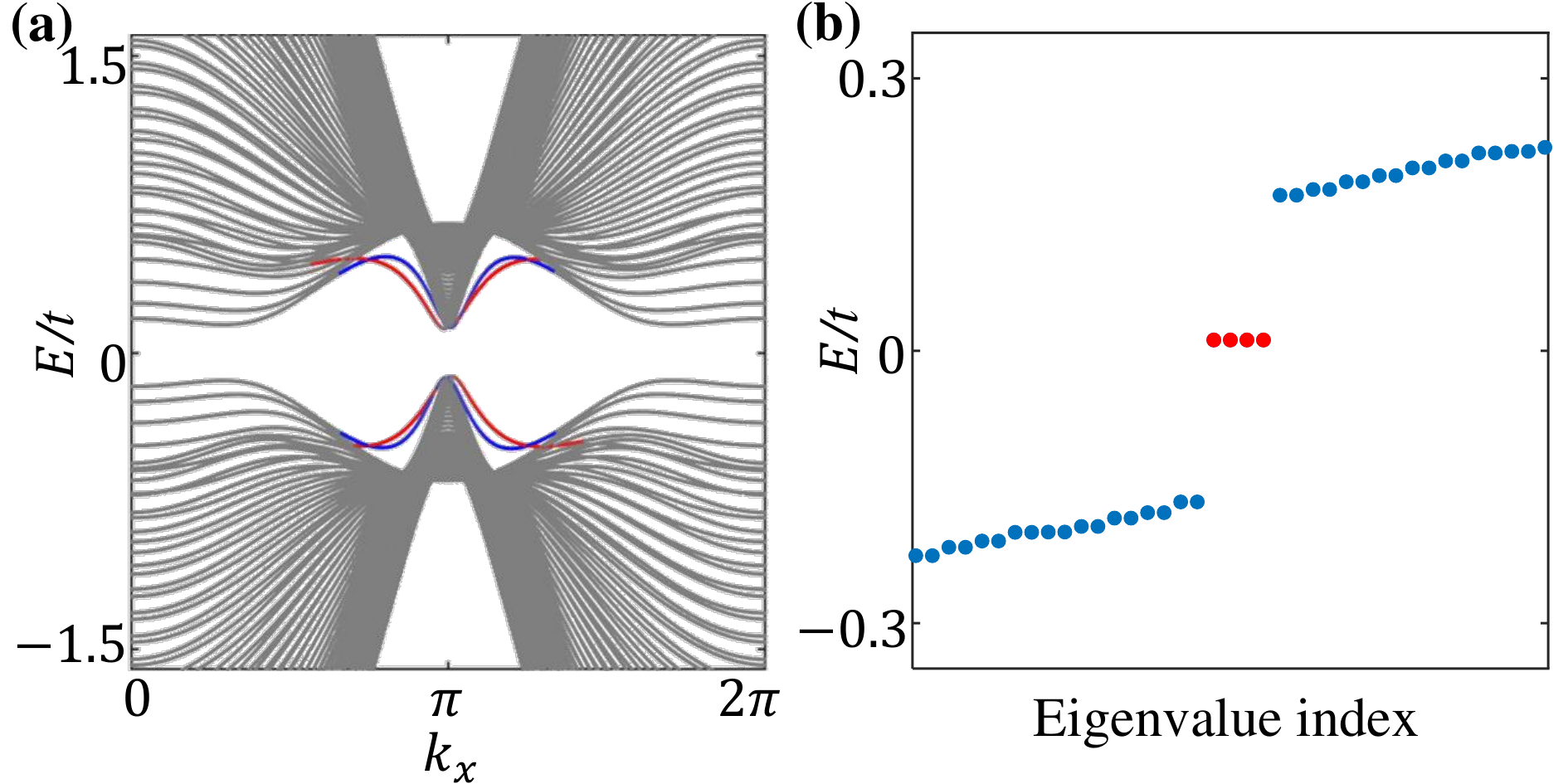}
    \caption{Robustness against screw-breaking perturbations. (a) Edge spectrum of a nanoribbon under the lattice-distortion perturbation $H_\text{d}$ ($\delta_1 = 0.02$). The red and blue lines denote the split, gapped edge states. (b) Energy spectrum of a fully finite sample under the same perturbation, where the four red dots highlight the robust corner states.}
    \label{fig4}
\end{figure}

The decoupling between the first- and second-order boundary sectors can be exposed explicitly by introducing the lattice-distortion perturbation
\begin{equation}
H_{\rm d}=\delta_1(\cos k_x-\cos k_y)\tau_z\sigma_0.
\label{eq:hpert}
\end{equation}
This perturbation breaks the two screw symmetries $\mathcal{S}_x$ and $\mathcal{S}_y$ and therefore immediately gaps out the first-order edge modes, as shown in Fig.~\ref{fig4}(a). At the same time, $H_{\rm d}$ preserves the magnetic mirrors $\mathcal{M}_x\mathcal{T}$ and $\mathcal{M}_y\mathcal{T}$, as well as the chiral symmetry generated by $\tau_y$. Consequently, the alternating edge-mass structure remains intact, and the four corner states survive in the fully finite geometry, as shown in Fig.~\ref{fig4}(b). This perturbative test clearly demonstrates that the first-order edge states and second-order corner states belong to distinct symmetry sectors of the same bulk AFM phase.

\section{Conclusion} We have established a symmetry-based framework for boundary-selective hybrid-order topology in two-dimensional nonsymmorphic antiferromagnets. A single AFM bulk insulating phase can display either first-order or second-order boundary topology depending solely on which crystalline symmetry sector is preserved by the physical termination. Screw-compatible edges host gapless Dirac modes, whereas screw-breaking diamond geometries support corner states generated by a magnetic-mirror-enforced inversion of the edge Dirac masses. The higher-order phase is further diagnosed by a quantized quadrupole moment, while the exact zero-energy pinning of the corner states is reinforced by the chiral symmetry of the minimal model.

These results show that the dimensionality of the topological boundary response in nonsymmorphic antiferromagnets can be controlled through boundary geometry without changing the underlying bulk phase. In this sense, the present model may be viewed as a symmetry-based insulating descendant of CuMnAs-type nonsymmorphic AFM Dirac platforms~\cite{wadley2013tetragonal,tang2016dirac,PhysRevLett.118.106402,Linn2023CuMnAs}, where a symmetry-allowed bulk mass gaps the Dirac sector and enables hybrid-order topology. More broadly, our work identifies a natural route toward boundary-engineered magnetic topological devices in which distinct topological responses are selected by crystallographic termination.

\begin{acknowledgments}
This work was supported by National Key Research and Development Program of the Ministry of Science and Technology of China (Grant No. 2025YFA1411303), the National Natural Science Foundation of China (NSFC, Grants No. 92565103, No. 12474151, No. 12222402, and No. 12547101), the Natural Science Foundation of Chongqing (Grant No. CSTB2025NSCQ-LZX0010), Beijing National Laboratory for Condensed Matter Physics (No. 2024BNLCMPKF025), and the Fundamental Research Funds for the Central Universities (Grant No. 2025CDJIAISYB-032).
\end{acknowledgments}

\bibliography{references}

\end{document}